# The influence of femtosecond laser wavelength on waveguide fabrication inside fused silica

Javier Hernandez-Rueda, Jasper Clarijs, Dries van Oosten, and Denise M. Krol

# The influence of femtosecond laser wavelength on waveguide fabrication inside fused silica

Javier Hernandez-Rueda,[1,2,a)] Jasper Clarijs,[2] Dries van Oosten,[2] and Denise M. Krol[1,a)]
[1]*Department of Materials Science and Engineering, University of California Davis, California 95616, USA*
[2]*Debye Institute for Nanomaterials Science, Utrecht University, P.O. Box 80000, 3508 TA Utrecht, The Netherlands*

We have investigated the effect of the laser wavelength on the fabrication of optical waveguides and tracks of modified material via direct laser writing inside fused silica. The size of the laser-inscribed tracks, the material modification thresholds, the structural changes, and the waveguide writing energy range show a strong dependence on laser wavelengths ranging from ultraviolet to near-infrared. We used numerical simulations that consider the laser-excited electron plasma dynamics (via multiple rate equations) along with Gaussian beams theory to calculate the size of the laser-affected volume that has been further compared with the experimental results. This study yields insight into how to predict and design the spatial features of laser-inscribed lines and also aids of understanding the underlying physical mechanisms linked to laser-glass interaction when using different laser wavelengths. *Published by AIP Publishing.*
[http://dx.doi.org/10.1063/1.4981124]

From the nineties, ultrashort laser pulses have been used for precise micromachining of dielectric materials due to the deterministic nature of the ionization mechanisms that mediate their energy coupling.[1,2] The extremely high power density achieved at the focal spot ($\sim 10$ TW/cm$^2$) triggers the generation of a dense electron plasma via strong field ionization (i.e., multi-photon and tunnel ionization) while the laser pulse is present ($\sim 100$ fs).[3,4] Afterwards, at the picosecond time-scale, the plasma energy is deposited into the glass network via electron-phonon interactions, among other mechanisms, which features a reduced heat affected zone.[5–8]

The advantages inherent to femtosecond (fs) laser pulses make them an ideal tool for fabricating optical waveguides inside glasses via direct laser writing.[9–12] The effect of the laser processing parameters such as pulse duration, polarization, pulse energy, writing speed, and repetition rate has been thoroughly investigated and extensively reported in the literature for a variety of glasses with different compositions.[12–16] Furthermore, the use of laser beam shaping techniques has been exploited (a) to find potential advantages for laser micromachining and (b) to study the influence of the temporal and spatial distribution of the laser on the balance of the ionization mechanisms.[17,18] However, the influence of the laser wavelength on the fabrication of optical waveguides has not been fully understood nor systematically investigated, although several commercially available laser wavelengths and their frequency doubled counterparts have been independently used for direct laser writing.[7–15] In this letter, the impact of the laser wavelength on the spatial features of laser-inscribed waveguides and tracks of damage in fused silica is studied. Laser waveguide writing experiments were performed with laser wavelengths from UV to near-infrared (NIR), attained using an optical parametric amplifier. A theoretical framework is presented using strong field ionization (SFI) and multiple rate equations (MRE) method.[19–22] The experimental results and their correspondence with numerical simulations are discussed.

Femtosecond waveguide writing in fused silica (7980 Corning) was carried out using laser wavelengths of 400, 800, 1100, 1200, 1300, and 1400 nm. The glass sample was polished to achieve high optical quality ($\sim \lambda/10$) by using an industrial polishing machine (Strasbaugh, Inc, 6DE-DC-1). The laser source used for the experiments was a regenerative amplifier (Spitfire LCX, Spectra-Physics). This system delivers ultrashort Fourier-limited pulses of 200 fs pulse duration at a wavelength of 800 nm with a repetition rate of 1 kHz. A commercial ultrafast OPA-800F (Spectra-Physics) was used for generating wavelengths between 1100 and 1400 nm. The 400 nm wavelength was obtained through second harmonic generation of the 800 nm beam in a BBO (beta-barium-borate) crystal.

The direct laser writing setup allows for diverse operation modes: laser longitudinal line inscription, optical microscopy, and waveguide near and far-field profile imaging. The laser beam passes through an energy control system ($\lambda/2$ wave-plate plus a polarizer) and through a confocal layout for mode cleaning (f = 500 mm, 150 $\mu$m aperture). The laser line inscription is then carried out by focusing the laser beam inside the glass sample (thickness = 2 cm), using an infinity corrected microscope objective (ELWD Nikon LU Plan, NA = 0.55, working distance = 10.1 mm), while it is translated at a constant speed (50 $\mu$m/s) using an air-bearing scan stage (AerotechFG1000). The laser longitudinal line inscription starts at the maximum depth allowed by the objective lens and continues until the focused beam is 300 $\mu$m from the sample surface; at that point the beam is blocked using a mechanical shutter in order to prevent surface ablation. Therefore, the input facet of a written-line is produced at a writing depth of 300 $\mu$m. An *in-situ* microscope

[a)]Authors to whom correspondence should be addressed. Electronic addresses: j.hernandezrueda@uu.nl and dmkrol@ucdavis.edu

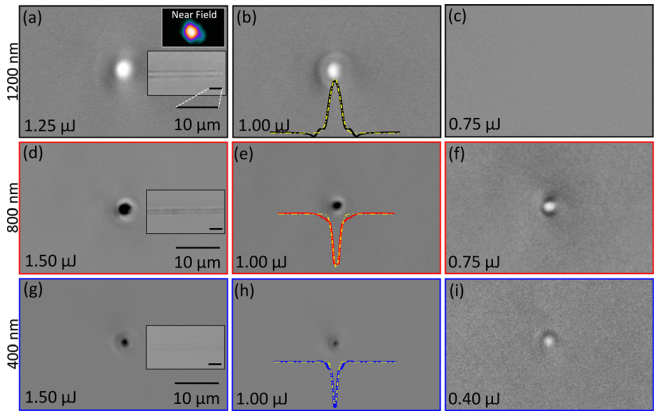

FIG. 1. (a)–(i) Set of white light micrographs along with averaged radial profiles (1 $\mu$J) of the frontal view of written lines inside fused silica using three laser wavelengths and pulse energies. The size brightness and contrast were set to be the same. The insets in (a), (d), and (g) show the lateral profiles. The yellow dotted lines are Gaussian fits to the profiles that determine the $1/e^2$ diameter. The near field profile at 660 nm ($9 \times 14$ $\mu m^2$) is shown as an inset to (a).

provides a first assessment of the laser-inscribed lines and their potential to operate as waveguides. The far and near field profiles were imaged with the same optical arrangement by coupling a CW-laser at 660 nm to the waveguides.[10,15]

Figure 1 shows a set of micrographs of the transverse cross section of the input facet of the inscribed lines for three representative wavelengths. The images feature three processing regimes, namely, unmodified material (Fig. 1(c)), smooth optical changes (Figs. 1(a), 1(b), 1(f), and 1(i)), and damaged material (Figs. 1(d), 1(e), 1(g), and 1(h)). Their circular appearance is due to the use of longitudinal writing, which was used in order to avoid the laser beam distortion and the loss of laser energy inherent to spherical aberration (SA) and slit shaping technique during perpendicular writing.[23] Waveguiding operation is experimentally found (and verified by measuring the near field, see Fig. 1(a)) for the energy range where smooth optical modifications are produced and positive refractive index change is achieved. The refractive index change of the waveguides was characterized by measuring the far field profiles of the guided modes at 660 nm. In this way, the diameter $D$ of the far field distribution (fitting the far field using a Gaussian function) was measured at a distance $L$ from the output facet of the waveguide (see Refs. 10 and 15 for a detailed explanation). The maximum refractive index increase ranges from $\Delta n/n \approx 1 \times 10^{-3}$ to $4 \times 10^{-4}$. The results are summarized in Table I. Here, three main observations can be extracted. First, the thresholds (both for damage and smooth changes) increase with the laser wavelength. Second, the lateral size of the machined lines decreases drastically when shorter laser wavelengths are utilized. Finally, for wavelengths up to 1100 nm, the range of energies where waveguides can be produced is broader for larger wavelengths also displaying a slightly increasing refractive index change.

The input facets of the inscribed lines were further inspected by using Raman confocal microscopy (CW-laser at $\lambda = 473$ nm, MO: NA = 0.5 and 100 $\mu$m pinhole)[24] in order to characterize the laser-induced structural modifications. Figure 2(a) illustrates the Raman spectrum of non-processed fused silica normalized to the 445 cm$^{-1}$ peak, where the Raman bands at 605 cm$^{-1}$ and 490 cm$^{-1}$ are, respectively, related to the breathing modes from 3- and 4-membered Si-O ring structures inside the glass network. The inset in Figure 2(b) shows a false color map of the relative increase of the 605 cm$^{-1}$ Raman band of the laser-modified cross-section shown in the optical micrograph on the left. Note that the $\mu$-Raman maps further validate the spatial dimensions of the lines that are obtained from the optical micrographs. The graph in Figure 2(b) presents the highest laser-induced structural change, of the 605 cm$^{-1}$ Raman band intensity with respect to the 445 cm$^{-1}$ peak ($\Delta I_{605}$), as a function of the laser energy for two representative wavelengths. The $\Delta I_{605}$ change increases with the laser energy, indicating an increase in the relative amount of 3-membered Si-O ring structures in the glass network of the laser processed region. This is in good agreement with previously reported results by our research group, where a growth of the area under the 605 cm$^{-1}$ band was also observed for increasing laser pulse energies for laser pulses at 800 nm.[24,25] As discussed in detail in previous work,[25,26] only moderate changes in the network structure (and thus in $\Delta I_{605}$) yield good waveguides, whereas more severe structural changes are associated with optical damage. For longer wavelengths, there is a wider range of laser pulse energies that result in moderate structural changes, consistent with the results reported in Table I.

In order to understand the observed behavior in terms of the transversal size of the written lines, we need to consider changes in the laser spot size as well as changes in the non-linear ionization process as a function of wavelength. The laser beam waist of a focused Gaussian beam is given by $w_o = \lambda/\pi NA$, where $\lambda$ is the wavelength and *NA* the numerical aperture of the focusing optics. The spatial intensity distribution of the laser is given by

$$I(r,z) = I_0 \left(\frac{w_o}{w}\right)^2 e^{-\frac{2(r-r_0)^2}{w^2}}; \quad w = w_o \sqrt{1 + \frac{z^2}{z_r^2}}, \quad (1)$$

where $I_o$ is the peak intensity and $z_r$ stands for the well-known Rayleigh range. Figure 3(a) illustrates 2D-maps of the spatial intensity distribution inside fused silica, calculated using representative experimental conditions, namely, focusing optics ($NA = 0.55$) and laser parameters ($I_o = 8 \times 10^{13}$ W/cm$^2$, $\lambda_1 = 1200$ nm, $\lambda_2 = 800$ nm, and $\lambda_3 = 400$ nm). The calculated maps show that the focal volume decreases for shorter wavelengths. Hence, for a given laser peak intensity, $I_o$, the alteration of the spatial distribution leads to different local intensities. This ultimately affects the way the laser energy is coupled through SFI and the extension of the laser-affected volume.

TABLE I. Experimental results.

| $\lambda$ (nm) | Energy range for waveguides ($\mu$J) | Diameter[b] of modification for 1 $\mu$J ($\mu$m) | $\Delta n/n_o$ at 1 $\mu$J | Maximum refractive index change $\Delta n/n_o$ |
|---|---|---|---|---|
| 400 | 0.20–0.40 | 2.1 | Damage | … |
| 800 | 0.30–0.75 | 3.4 | Damage | $(4.2 \pm 0.9) \times 10^{-4}$ |
| 1100 | 0.75–3.00 | 4.3 | $4.4 \times 10^{-4}$ | $(6.6 \pm 1.7) \times 10^{-4}$ |
| 1200 | 1.00–1.75[a] | 4.6 | $2.4 \times 10^{-4}$ | $(6.8 \pm 1.7) \times 10^{-4}$ |
| 1300 | 1.25–1.60[a] | …[c] | …[c] | $(16.5 \pm 4.1) \times 10^{-4}$ |

[a]Available laser pulse energy was not sufficient to produce damage or higher refractive index change.
[b]The experimental $1/e^2$ radius was retrieved by fitting the averaged radial profiles in Fig. 1 to a Gaussian function.
[c]No laser modification for 1300 nm at 1 $\mu$J.

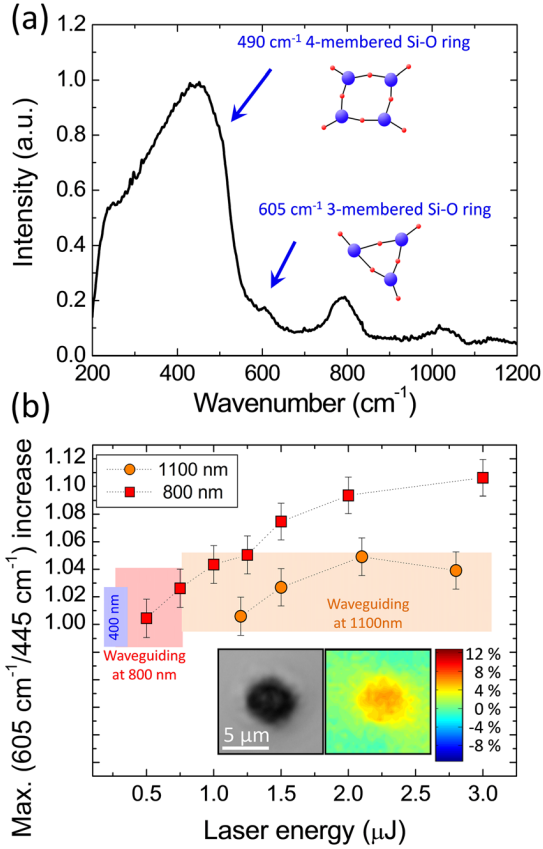

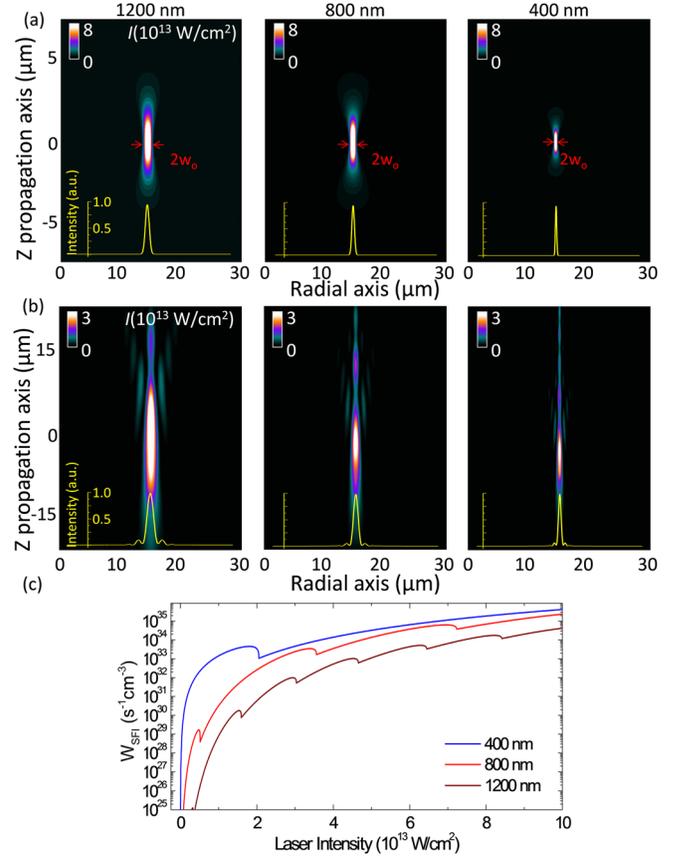

FIG. 2. (a) Raman spectrum for unmodified fused silica normalized to the 445 cm$^{-1}$ peak. (b) Relative increase of the Raman peak intensity $\Delta$(605 cm$^{-1}$/445 cm$^{-1}$) as a function of the laser energy. The inset shows an optical micrograph and a false color map of the 605 cm$^{-1}$/445 cm$^{-1}$ Raman peak intensity for a line written using 800 nm. The colored rectangles in (b) show the region where waveguide operation is achieved.

FIG. 3. False color maps, for three laser wavelengths, of (a) the laser intensity distribution $I(r,z)$ of a tightly focused ($NA = 0.55$) Gaussian beam. The maps in (b) illustrate the laser intensity distribution considering the effect of SA when focusing the beam 300 $\mu$m under the sample surface. The insets in (a) and (b) show the cross-sections (yellow lines) of the normalized intensity distributions where the peak power is achieved within the maps. (c) SFI rate, $W_{SFI}(\lambda,I)$, is plotted as a function of the laser intensity.

The maps in Fig. 3(b) present the intensity distribution of a tightly focused Gaussian beam in the presence of a refractive index mismatch (air-glass interface), which generates SA. The electric field distribution is calculated using the diffraction integral derived by Török *et al.* for a Gaussian laser beam.[27] For the calculation, we have used the focusing depth (300 $\mu$m) at which the experimental diameters of the laser-inscribed lines are characterized in Fig. 1. These maps show how the focal volume suffers an elongation along the laser propagation axis (also the writing direction) due to SA. The insets in Figs. 3(a) and 3(b) present the lateral cross-section (yellow lines) of the laser intensity for the z position where the peak intensity is achieved. The lateral distributions show a moderate increase, which also causes a decrease in the maximum peak power achieved. Therefore, the intensity distributions affected by SA will be used for the simulation of the transversal size of the laser-inscribed lines. The values of the 1/e$^2$ beam waist considering the effect of SA, $w_o(SA)$ are shown in Table II.

To calculate the size of the laser-modified volume, it is necessary to set a criterion for modification. In this work, we hypothesize that the material will be modified when the laser-generated electron density exceeds a specific value, $n_e$ ($\sim 10^{21}$ cm$^{-3}$).[16,28–32] Electrons are initially excited through nonlinear photo-ionization. The photo-ionization rate $W_{SFI}$ can be calculated using the Keldysh formalism[19,20,33]

$$W_{SFI}(\omega_L, U_g, E_L) = 2\frac{2\omega_L}{9\pi}\left(\frac{m\omega_L}{\hbar\sqrt{\gamma_1}}\right)^{\frac{3}{2}} Q(\gamma,x) e^{-\pi\langle x+1\rangle \frac{K_{\gamma_1}-E_{\gamma_1}}{E_{\gamma_2}}}, \quad (2)$$

$$\gamma = \frac{\omega_L\sqrt{mU_g}}{eE_L}, \quad \gamma_1 = \frac{\gamma^2}{1+\gamma^2}, \quad \gamma_2 = 1 - \gamma_1,$$

$$x = \left(\frac{2U_g}{\pi\omega_L\sqrt{\gamma_1}}\right) E_{\gamma_2}, \quad (3)$$

TABLE II. MRE simulation results ($I$ and $d$), $W_{1pt}$ coefficients, and $w_o$ considering SA.

| $\lambda$ (nm) | Intensity for $n_e = 10^{21}$ cm$^{-3}$ ($10^{13}$ W/cm$^2$) | Diameter of modification ($\mu$m) | $W_{1pt}$[a] ($10^{-6}$ $E_L^2$ m$^2$/V$^2$s) | $w_o$(SA) ($\mu$m) |
|---|---|---|---|---|
| 400 | 1.70 | 0.972 | 0.195 | 0.47 |
| 800 | 2.44 | 1.476 | 1.380 | 0.79 |
| 1100 | 2.30 | 1.764 | 3.158 | 1.04 |
| 1200 | 2.35 | 1.836 | 3.911 | 1.06 |

[a]$W_{1pt}$ was calculated as in Refs. 17 and 22.

$$Q(\gamma, x) = \sqrt{\frac{\pi}{2K_{\gamma_2}}} \sum_{n=0}^{\infty} \exp\left(-n\pi \frac{K_{\gamma_1} - E_{\gamma_1}}{E_{\gamma_2}}\right)$$
$$\times \Phi\left(\sqrt{\frac{\pi^2(\lfloor x+1 \rfloor - x + n)}{2K_{\gamma_2} E_{\gamma_2}}}\right), \quad (4)$$

$$U_g = U_g^0 + \frac{e^2 E_L^2}{4m\omega_L^2}, \quad (5)$$

where $\omega_L$ stands for the laser frequency, $U_g$ for the bandgap ($U_{g0} = 9$ eV for fused silica), $E_L$ is the electric field strength of the laser, $K_{\gamma i}$ and $E_{\gamma i}$ denote the complete elliptic integral of the first and second kind, $m$ and $e$ are the reduced mass and the charge of the electron, $\gamma$ is the Keldysh parameter, and $\Phi$ is the Dawson integral. Figure 3(c) shows the ionization rate ($W_{SFI}$) curves as a function of the laser intensity calculated by using formulas (2)–(5). For a given intensity value, the graphs illustrate significantly higher peak ionization rates for shorter laser wavelengths.

However, SFI is not the only process by which the laser energy is absorbed or free carriers are created. An electron, excited via SFI, can increase its energy through a succession of single-photon absorption events (inverse bremsstrahlung). When the electron energy reaches an energy of $U_g$, it can ionize an electron in the valence band via impact ionization. The number of photons that the avalanche ionization (AI) process requires is given by $k = \lceil U_g/\hbar\omega_L \rceil$. Note that $k$ is thus also wavelength dependent. We have calculated the total electron density using Rethfeld's multiple rate equation (MRE) method.[22] In this method, the conduction band is modeled as a set of discrete energy levels with an energy spacing equal to the photon energy $\hbar\omega_L$. Initially, the lowest level is populated by SFI, and subsequent levels are coupled by one photon absorption. The population in the highest level is coupled to the lowest level describing the process of impact ionization. The equations for the electron populations in the conduction band are given by

$$\begin{aligned}\dot{n}_0 &= W_{SFI} + 2\alpha n_k - W_{1pt}(E_L)n_0 \\ \dot{n}_1 &= W_{1pt}(E_L)n_0 - W_{1pt}(E_L)n_1 \\ &\vdots \\ \dot{n}_{k-1} &= W_{1pt}(E_L)n_{k-2} - W_{1pt}(E_L)n_{k-1} \\ \dot{n}_k &= W_{1pt}(E_L)n_{k-1} - \alpha n_k,\end{aligned} \quad (6)$$

where $\alpha$ is the avalanche ionization coefficient and $W_{1pt}$ is the one photon absorption coefficient (Table II). The density required for producing smooth optical modifications inside fused silica has been reported to be $n_e \approx 10^{21}$ cm$^{-3}$.[16,28–32] In this letter, we target such density value in order to find the required input laser intensity. The wavelength dependent rates $W_{SFI}(\lambda)$[33] and $W_{1pt}(\lambda)$[17,22] and the material properties of fused silica ($U_{g0} = 9$ eV, $n(\lambda)$) were used for these simulations. The SFI rate curves were calculated by using the Keldysh formalism. The maximum number of equations, $k + 1$, is calculated for the peak electric field of the laser. However, $U_g$ is updated dynamically. Thus, for the simulations, the conduction band level in which the avalanche ionization coefficient ($\alpha = 1$ fs$^{-1}$) is located at certain time iteration is shifted depending on $U_g$ ($E_L(t)$).

Figure 4 shows the electron density as a function of time, where the same final target density ($10^{21}$ cm$^{-3}$) is achieved using different laser intensities for each wavelength (see Table II). The influence of the wavelength on the electron density, through $W_{SFI}(\lambda)$ and $W_{Ipt}(\lambda)$, can be appreciated in the balance of the ionization mechanisms. Although the four curves achieve the same maximum $n_e$, the relative AI contribution for longer wavelengths (i.e., 1200 nm) is considerably higher (MRE patterned grey area) than for their shorter counterparts. This type of ionization-mechanism-detuning effect and its influence on the permanent material modifications have been analogously observed using temporal pulse shaping for NIR wavelengths.[17,34–36]

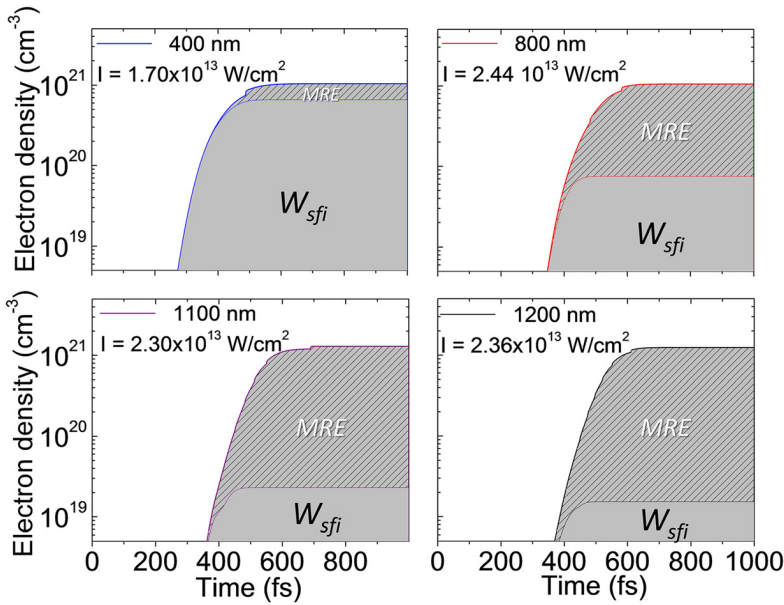

FIG. 4. Electron density as a function of time for different laser wavelengths and intensities. Higher and lower $n_e$ correspond to the electron density, respectively, calculated using MRE and SFI-only.

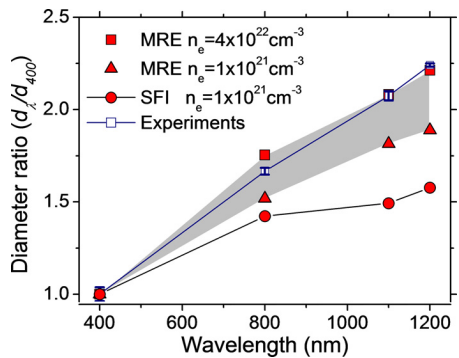

FIG. 5. Diameter ratio ($d_\lambda/d_{400\,nm}$) of the laser-affected focal volume inside fused silica calculated considering SA. Red and white symbols are, respectively, linked to theoretical and experimental results. The grey area is limited by the diameters estimated for electron plasma densities ranging from $1 \times 10^{21}$ cm$^{-3}$ to $4 \times 10^{22}$ cm$^{-3}$.

Our calculations yield values for the local intensities that are needed to reach the target electron density. We also know the laser intensity distributions $I(r,z,\lambda)$ for the actual experimental parameters (1 µJ, NA = 0.55, $\lambda$ = 400, 800, 1100, 1200 nm) calculated considering the experimental writing depth (300 µm).[27] From these distributions, we can determine the diameter of the area where the target e$^-$ density is exceeded (see Table II). To focus on the wavelength dependence rather than the absolute values, the graph in Figure 5 presents the diameter ratio with respect to the diameter at 400 nm ($d_\lambda/d_{400\,nm}$). The grey area accounts for the range of diameter ratios achieved when the density of the electron plasma reaches values between $1 \times 10^{21}$ and $4 \times 10^{22}$ cm$^{-3}$, which have been calculated using the MRE method and considering the effect of SA. The red circles underneath indicate ratios where the calculations are carried out using SFI-only. It illustrates the necessity to use a description of the laser energy deposition where AI, SFI, and inverse bremsstrahlung are taken into account. Overall, the experimentally measured $d_\lambda/d_{400\,nm}$ trend shows good agreement with the numerical simulations obtained using MRE and considering the impact of the writing depth.

In conclusion, this letter investigates the role of the laser wavelength during direct laser writing inside fused silica. The fabrication of waveguides and tracks of damage have been demonstrated using laser wavelengths from UV to NIR. Shorter wavelengths lead to smaller laser-written spatial features, lower thresholds, and a narrower energy range for fabricating waveguides. The simulation of the electron plasma density dynamics, via MRE, exhibits a lower implication of SFI compared with AI and inverse bremsstrahlung when using NIR laser wavelengths. The experimentally and numerically attained diameters of the laser-affected zone have been compared, showing excellent agreement.

This material is based upon work supported by the National Science Foundation under Grant No. DMR 1206979. J.H.R. acknowledges support through Marie Skłodowska-Curie Action IF (703696/ADMEP).